\documentclass[acmsmall]{acmart}
\usepackage{caption}
\usepackage{subcaption}
\usepackage{multirow}
\usepackage{tabularx}
\usepackage{makecell}
\usepackage{booktabs}
\usepackage{url}

\AtBeginDocument{%
  }

\setcopyright{cc}
\setcctype{by}
\acmDOI{10.1145/3797121}
\acmYear{2026}
\acmJournal{PACMSE}
\acmVolume{3}
\acmNumber{FSE}
\acmArticle{FSE026}
\acmMonth{7}
\acmSubmissionID{fse26maina-p500-p}
\received{2025-09-11}
\received[accepted]{2025-12-22}

\begin{document}

\title{Recommending Usability Improvements with Multimodal Large Language Models}

\author{Sebastian Lubos}
\orcid{0000-0002-5024-3786}
\affiliation{%
  \institution{Graz University of Technology}
  \city{Graz}
  \country{Austria}
}
\email{sebastian.lubos@tugraz.at}

\author{Alexander Felfernig}
\orcid{0000-0003-0108-3146}
\affiliation{%
  \institution{Graz University of Technology}
  \city{Graz}
  \country{Austria}
}
\email{alexander.felfernig@tugraz.at}

\author{Damian Garber}
\orcid{0009-0005-0993-0911}
\affiliation{%
  \institution{Graz University of Technology}
  \city{Graz}
  \country{Austria}
}
\email{damian.garber@tugraz.at}

\author{Viet-Man Le}
\orcid{0000-0001-5778-975X}
\affiliation{%
  \institution{Graz University of Technology}
  \city{Graz}
  \country{Austria}
}
\email{v.m.le@tugraz.at}

\author{Manuel Henrich}
\orcid{0009-0007-4644-0836}
\affiliation{%
  \institution{UNiQUARE Software Development}
  \city{Krumpendorf am Wörthersee}
  \country{Austria}
}
\email{manuel.henrich@uniquare.com}


\begin{abstract}
\textit{Usability} describes quality attributes of application \textit{user interfaces} that determine how effectively users can interact with them. Traditional usability evaluation methods require considerable expertise and resources, which can be challenging, especially for small teams and organizations. Automating usability evaluation could make it more accessible and help to improve the user experience.
The recent emergence of powerful \textit{multimodal large language models (MLLMs)} has opened new opportunities for automating usability evaluation and recommendation of improvements. These models can process visual inputs such as images and videos alongside textual context, which enables the identification of usability issues and the generation of actionable suggestions to resolve these issues. 

In this paper, we present a novel automated approach that uses limited application context and screen recordings of user interactions as input to an MLLM. The model automatically identifies and describes usability issues based on \textit{Nielsen’s usability heuristics}, and provides corresponding explanations and improvement recommendations. To reduce the developer effort of manual prioritization, the recommendations are ranked by severity.
The quality and practical usefulness of the generated recommendations were evaluated based on a user study that involved software engineers as participants. The evaluation focused on the highest-ranked suggestions provided by the model. The results demonstrate the potential of our approach to provide low-effort usability improvement recommendations. This makes it a promising complement to traditional evaluation methods, especially in settings with limited access to usability experts. In this sense, the approach serves as a basis for future integration into development tools to enable automated usability evaluation within software engineering workflows.
\end{abstract}




\maketitle

\section{Introduction}\label{sec:introduction}

The \textit{usability} of interactive software systems describes quality attributes that reflect how effectively and efficiently users can complete their intended tasks while using the system~\cite{winter2008comprehensive}. Good usability can help with the acceptance and usage of a system, while poor usability may lead to lower user satisfaction, reduced adoption, and higher costs for support and maintenance~\cite{nielsen2012introduction}. Though usability is important and should be evaluated throughout the entire \textit{software development lifecycle} \cite{ruparelia2010software}, it is often out of focus, particularly in small teams or companies that lack expertise or resources.

Traditional usability evaluation methods include \textit{usability testing}, where real users are observed while solving tasks~\cite{hass2019practical}, and \textit{usability inspection}~\cite{holingsed2007usability}, where experts simulate the task-solving of real users. Both aim to identify usability issues but are resource-intensive, as they require manual analysis, expert knowledge, and dedicated time. This makes it difficult to apply them regularly and integrate them into continuous development workflows~\cite{carvajal2013usability}. To support developers with limited experience, heuristic guidelines for systematic evaluation, such as the \textit{Nielsen heuristics}~\cite{nielsen1994enhancing}, have been proposed. Yet, they do not replace the effort of manual evaluations. Automation can reduce this effort and help to identify usability issues early on in the software development lifecycle~\cite{ruparelia2010software}.

Although several automated evaluation tools have been suggested for this purpose~\cite{namoun2021review,castro2022automated,kuric2025systematic}, there is still no solution that supports all usability-relevant dimensions. Recent versions of \textit{multimodal large language models (MLLMs)}~\cite{yin2024survey} offer new opportunities for the automation of software engineering processes and the enhancement of software quality. MLLMs are capable of processing both textual and visual inputs, including images and videos. In this sense, they can help to automate usability evaluation by interpreting UI behavior and context from provided screenshots or screen recordings. Initial studies report promising results in identifying usability issues in design mockups~\cite{duan2024generating} and mobile applications~\cite{pourasad2025does}. While this automated evaluation could not yet replace expert validation, they were considered a helpful supplement to test usability, especially for small teams with limited resources. Further studies revealed that LLM-based analysis often aligned with expert assessments~\cite{zhong2025synthetic,lubos2026investigating}. This indicates that MLLMs could at least partly automate usability evaluation and support software developers, especially if they have limited usability expertise, to effectively shift usability assessments to earlier stages of the software development lifecycle.

Building on these encouraging results, we present a novel developer-centric approach that uses multimodal LLMs for automated, heuristic-based usability evaluation, which involves:

\begin{itemize}
    \item \textbf{Dynamic Interaction Analysis:} Unlike prior work using static screenshots, our approach uses video screen recordings to capture the dynamic nature of user interactions. This enables the identification of temporal usability issues, e.g., missing load indicators after a click. Using videos resembles real-life evaluation scenarios, similar to those performed by human experts.
    \item \textbf{Heuristic-Guided Evaluation:} We instruct the MLLM to evaluate UI behavior explicitly using Nielsen's heuristics~\cite{nielsen1994enhancing}, which are commonly used by human experts. This guides the model toward a structured assessment with a broad and complete evaluation of aspects.
    \item \textbf{Actionable Recommendations:} Beyond documenting identified usability issues, our approach suggests explicit solutions to resolve them and improve the usability. These recommendations are ranked by severity to reduce the need to manually prioritize the order of resolving identified issues and enhance the practical value for development teams.
\end{itemize}

To evaluate our approach, we conducted a user study with participants who have academic or professional experience in software engineering. They reviewed the usability issues identified by the MLLM and its improvement recommendations for two applications in terms of clarity, plausibility, and completeness. The results indicate that software engineers considered the outputs useful and applicable, which demonstrates the potential of MLLM-based analysis as a scalable automation technique to complement established usability evaluation practices. Rather than aiming to replace usability experts, this work focuses on supporting software engineers who lack extensive usability expertise. Overall, this work serves as a basis for the integration of automated usability evaluation in software developer workflows, including testing pipelines, continuous evaluation environments, and design tools.

The main contributions of this paper are the following:
\begin{enumerate}
    \item A novel automation technique for automated heuristic-guided usability evaluation using multimodal LLMs and screen recordings to consider dynamic user interactions.
    \item An automated developer-centric approach for generating and ranking actionable usability improvement suggestions by severity to reduce the need for manual prioritization.
    \item An empirical evaluation of the perceived quality and usefulness of the improvement recommendations through a user study with experienced software engineers.
\end{enumerate}

The remainder of this paper is organized as follows: Section~\ref{sec:automated} describes the details of our automated usability evaluation approach. Section~\ref{sec:prompt_engineering} describes the LLM configuration and prompt engineering. The study design is outlined in Section \ref{sec:study_design}. Section~\ref{sec:results} presents the study results. Section~\ref{sec:discussion} discusses practical implications. Section~\ref{sec:threats} addresses threats to validity. Section \ref{sec:related_work} summarizes related work. Finally, the paper is concluded in Section \ref{sec:conclusion}.

\section{Automated Usability Improvement Recommendation}\label{sec:automated}

\subsection{Overview}
\begin{figure}
    \centering
    \includegraphics[width=1.0\linewidth]{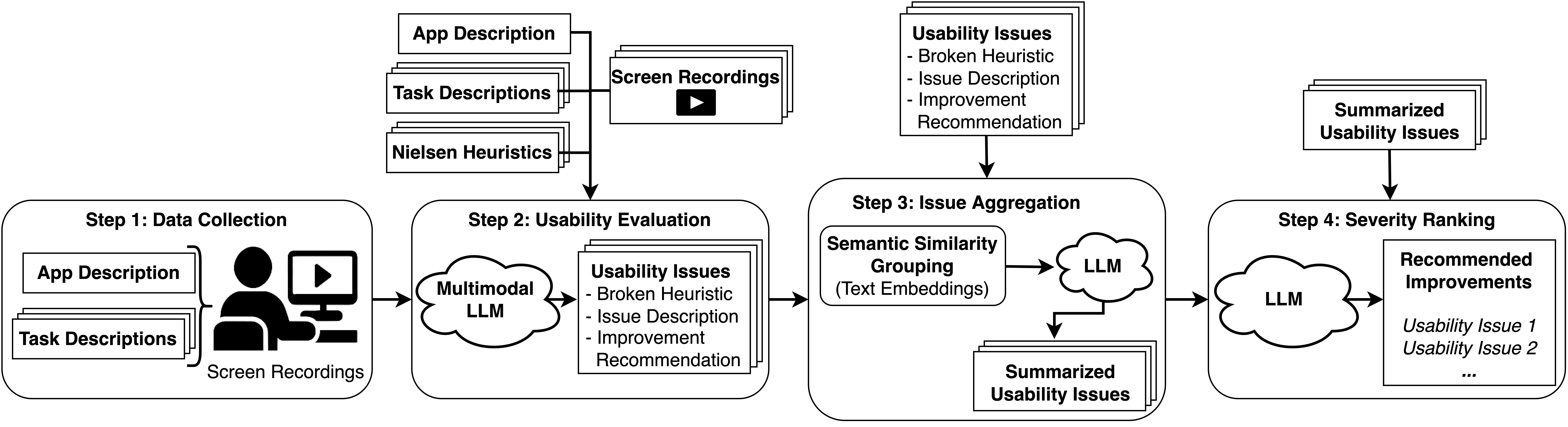}
    \caption{Overview of the four-step process for our automated heuristic-based usability evaluation and improvement recommendation. In step one, application data is collected by recording users while solving specified tasks. In step two, a multimodal LLM analyzes the interactions using Nielsen’s heuristics to identify usability issues and suggest improvements. Step three identifies and summarizes similar issues to reduce redundancy. Finally, the usability issues are ranked by severity to help prioritize improvements.}
    \label{fig:overview}
\end{figure}

Our approach includes four steps for the heuristic-based usability evaluation (see Figure \ref{fig:overview}):

\begin{enumerate}
    \item \textbf{Data Collection:}  
    A general description of the application and specific task instructions are manually created to prepare evaluation scenarios in line with traditional usability evaluation practices. Based on these instructions, a user performs the tasks individually using the application while their screen interactions are recorded in a video. A separate video is recorded for each task.

    \item \textbf{Usability Evaluation:}  
    The screen recording of a user performing a task, the app description, and the task description are provided as contextual input to a multimodal LLM. The LLM is instructed to analyze the application based on Nielsen’s usability heuristics. For each heuristic, the model generates a description of identified usability issues and recommends actionable suggestions for resolving them.

    \item \textbf{Issue Aggregation:}  
    To reduce redundancy among issue descriptions that may have been identified for multiple heuristics, similar issues are identified and aggregated. Semantic similarity measures are used to group similar issue descriptions and recommendations, which are then summarized by an LLM. This summary preserves heuristic-specific nuances and avoids repetition.

    \item \textbf{Severity Ranking:}  
    The aggregated usability issues are ranked by severity. The list of identified issues is provided as context to an LLM, which is instructed to evaluate the impact of each issue and order them by severity. This prioritization helps developers to focus on the most critical improvements first.
\end{enumerate}

This structured process enables scalable, automated usability evaluation by identifying issues and providing prioritized, actionable recommendations.

\subsection{Input Context}
To perform the automated evaluation, contextual input of the application to be evaluated is provided to the MLLM. This input includes a general textual description of the application to explain the purpose for which it is used. For the \textit{KnowledgeCheckR} application used in our evaluation (see Section \ref{ssec:application_context}), a description as follows is provided: ``\textit{KnowledgeCheckR is a browser-based learning platform offering microlearning, spaced repetition, and gamification. It supports multiple question types, provides instant feedback, and is used for training, exams, and compliance in corporate and educational settings.}'' This explanation gives a general idea of the app's purpose to set the context.

Additionally, different task descriptions are provided to explain the user's intention in the specific scenario. This consists of two parts. Firstly, a basic description of the assumed persona (i.e., the role of the user), e.g., ``\textit{A teacher who wants to create and manage knowledge checks (quizzes) to support student learning effectively.}'' Secondly, a concrete task to be solved, e.g., ``\textit{Create a new knowledge check by providing a title, image, description, deadline, and adding 2 multiple-choice questions. Complete all required fields and save the knowledge check.}'' These descriptions are used in the data collection phase to record the screen interactions, as well as in the usability evaluation with the MLLM, to set the focus of the evaluation.

The screen recordings are the most critical input required for the automated usability evaluation. They capture the temporal dynamics of the application behavior during user interaction and provide the visual context for the evaluation. In contrast to using simpler screenshots, this allows the evaluation to recognize issues related to the dynamic flow, e.g., missing load indicators after clicks. At the same time, it avoids the necessity to manually identify representative screenshots. The audio track of videos is not used in this version.

\subsection{Heuristic-Guided Evaluation} \label{ssec:heuristic}
To resemble an expert-like usability evaluation, the MLLM is explicitly instructed to assess the interface according to Nielsen’s ten usability heuristics,\footnote{\url{https://www.nngroup.com/articles/ten-usability-heuristics/}} which were selected for their general applicability across domains and broad adoption in usability practice and research. For each heuristic, the MLLM considers the provided screen recording, app description, and task instruction to identify potential violations. For identified issues, the model generates a concise description of the problem and an actionable recommendation for resolving it. This structured heuristic-based evaluation aligns with established inspection methods and ensures that key usability dimensions are covered. While Nielsen’s heuristics provide a suitable baseline, the approach could be extended with domain-specific or accessibility-focused heuristics to better address specialized contexts.


To ensure that the LLM uses the desired interpretation of heuristics, an explicit question is provided in the instructions to explain it. For example, while analyzing the \textit{User Control and Freedom} heuristic, the LLM instruction includes the description to focus on the question: ``\textit{Can users easily undo or exit unwanted actions (e.g., cancel, back, undo)?}''

While it would be possible to evaluate all heuristics at once, our approach was designed to consider each heuristic individually. This helped to obtain more diverse and detailed issue descriptions. Nevertheless, this also leads to redundancies in the list of identified issues, as the same issue can be related to multiple heuristics. To avoid returning these redundancies to the user, but at the same time keep the nuanced and detailed evaluation, we included an intermediate step for issue aggregation before ranking the issues by severity.

For this purpose, a \textit{semantic similarity} analysis was implemented. Semantic similarity compares the meanings of texts using textual embeddings \cite{patil2023survey}, which are high-dimensional vector representations of texts. Texts with similar meanings are closer in the vector space. We compute \textit{cosine similarity} between the embeddings to identify and group similar issues~\cite{li2013distance}. Higher values indicate greater similarity.

A text embedding model was used to compute the vector representations of the generated issue descriptions.\footnote{We used \textit{all-MiniLM-L6-v2} (see \url{https://huggingface.co/sentence-transformers/all-MiniLM-L6-v2}) as a practical design choice. Systematic comparison of embedding models was infeasible within the scope of a user study and remains an opportunity for future work.} If the cosine similarity of two or more description embeddings surpassed a threshold, they were grouped, as we assume they describe the same issue. After all groups have been identified, an LLM is instructed to aggregate the issue descriptions and related improvement recommendations by summarizing them to remove redundancies, while still respecting the nuanced aspects of the individual descriptions in the summaries. The threshold parameter can influence the quality of the final set of identified issues. A good trade-off between separating partly distinct issues and avoiding many redundancies is needed. We determined $0.7$ as a feasible value in our experiments. As the descriptions were summarized by an LLM, we assumed that even if some issues were wrongly grouped, the individual key aspects were still represented in the aggregated description.

\subsection{Severity Ranking} \label{ssec:severity}
As the number of identified usability issues can be high and cause considerable effort for manual review, our approach includes a final step to rank the issues by severity. This way, developers can prioritize and review the most critical issues first to reduce manual effort. For this purpose, the list of summarized usability issues per task and screen recording is provided as input for an LLM, which ranks them by severity. The model was instructed to consider severity-related aspects like impact on task success, user frustration, likelihood of occurrence, and effort for recovery. The usability issues were ranked individually for each investigated user task scenario.

We considered the possibility of retrieving an explicit severity rating by the LLM during the evaluation as an alternative to the separate rating step. However, this led to multiple identical ratings in our experiments, which made it harder to achieve a strict order. The observations were similar for using a 3-point rating scale (\textit{no issue}, \textit{minor issue}, or \textit{major issue}), or a 5-point scale with school grades. Therefore, we decided to use the approach based on relative ordering.

As a final output, the user receives a ranked list of usability issues, including a description of the identified issues and practical recommendations on how to resolve them. The violated usability heuristics per issue are provided as additional information. Based on this list, users can decide if and when they want to resolve the identified issues.

\section{Prompt Engineering and Model Setup}\label{sec:prompt_engineering}

\subsection{Prompt Design}
Our automated heuristic-based usability evaluation approach uses three dedicated prompt templates for its sub-tasks: (1) \textit{Usability Evaluation}, (2) \textit{Issue Aggregation}, and (3) \textit{Severity Ranking}. Each prompt was designed to explicitly define the model’s role and provide structured input to reduce ambiguity and minimize hallucinations. A dedicated system prompt is used to establish the intended LLM behavior and ensure a consistent evaluation strategy. Formatting instructions are included to return JSON output for further processing.\footnote{We tested different prompts informally and selected the final templates as a practical design choice. Systematic comparison of alternative prompt designs was infeasible within the scope of a user study and remains an opportunity for future research.}

\subsubsection{Usability Evaluation}
\begin{figure}[h]
    \centering
    \includegraphics[width=0.9\linewidth]{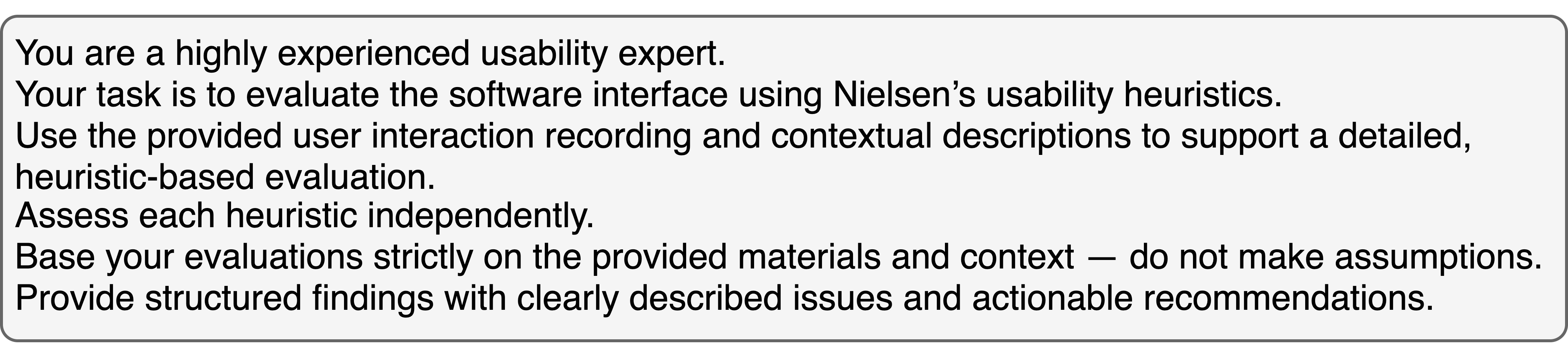}
    \caption{System prompt for usability evaluation.}
    \label{fig:system_usability_evaluation}
\end{figure}

\begin{figure}[h]
    \centering
    \includegraphics[width=0.9\linewidth]{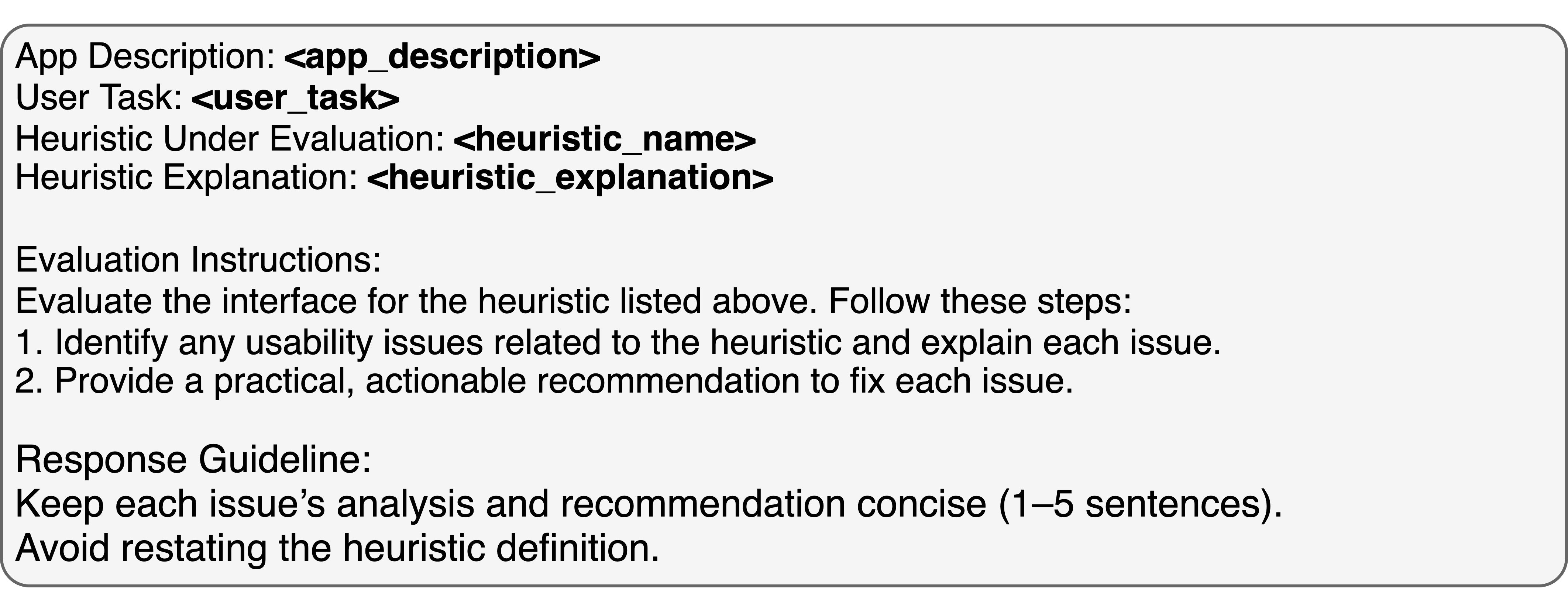}
    \caption{Prompt template for usability evaluation with variable placeholders in parentheses ($<>$). Result formatting instructions were omitted for brevity.}
    \label{fig:usability_evaluation}
\end{figure}

Figure~\ref{fig:system_usability_evaluation} shows the system prompt used for the LLM to evaluate the applications' usability based on Nielsen’s heuristics. It defines the model's role as a usability expert conducting a heuristic inspection. The user prompt in Figure~\ref{fig:usability_evaluation} provides detailed context, including the app description, task instructions, and heuristic to be evaluated.

\subsubsection{Issue Aggregation}
\begin{figure}[h]
    \centering
    \includegraphics[width=0.9\linewidth]{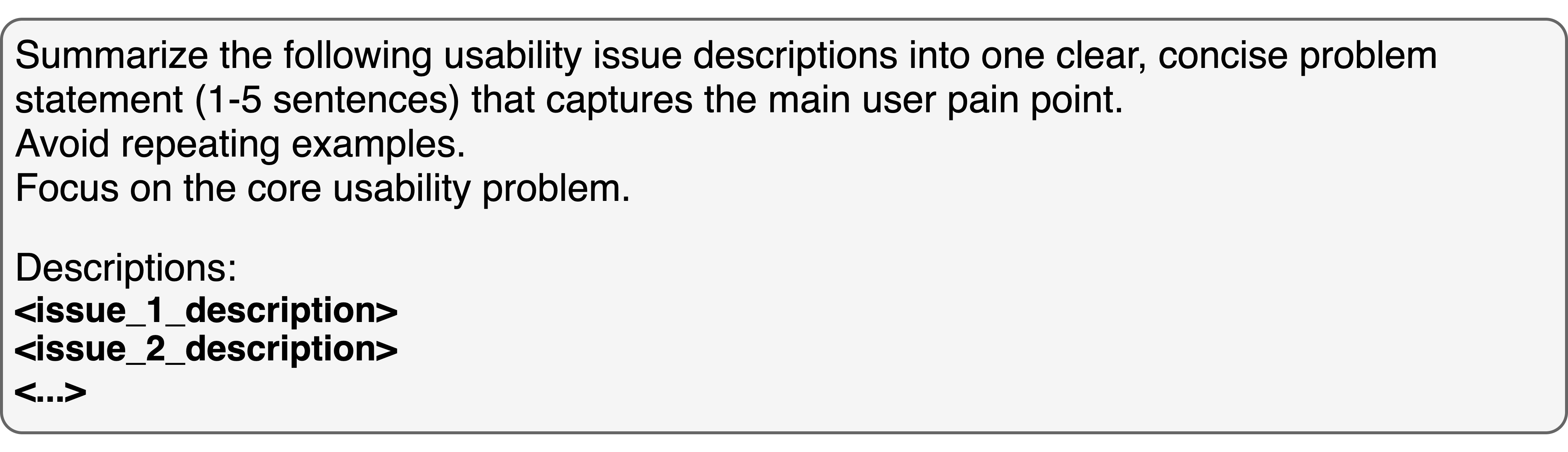}
    \caption{Prompt template for usability issue summary with variable placeholders in parentheses ($<>$). Result formatting instructions were omitted for brevity.}
    \label{fig:issue_aggregation}
\end{figure}

To aggregate the descriptions and improvement recommendations of similar usability issues (see Section~\ref{ssec:heuristic}), each group of similar issues is passed to the LLM using the prompt templates shown in Figures~\ref{fig:issue_aggregation} and~\ref{fig:issue_recommendation_aggregation}. These prompts instruct the model to generate a concise summary of the grouped issues and a unified improvement recommendation. Any heuristic-specific nuances that may add context or detail should be preserved to ensure that the output remains comprehensive yet non-redundant.

\begin{figure}[h]
    \centering
    \includegraphics[width=0.9\linewidth]{images/issue_improvement_aggregation.png}
    \caption{Prompt template for usability improvement recommendation summary with variable placeholders in parentheses ($<>$). Result formatting instructions were omitted for brevity.}
    \label{fig:issue_recommendation_aggregation}
\end{figure}

\subsubsection{Severity Ranking}
\begin{figure}[h]
    \centering
    \includegraphics[width=0.9\linewidth]{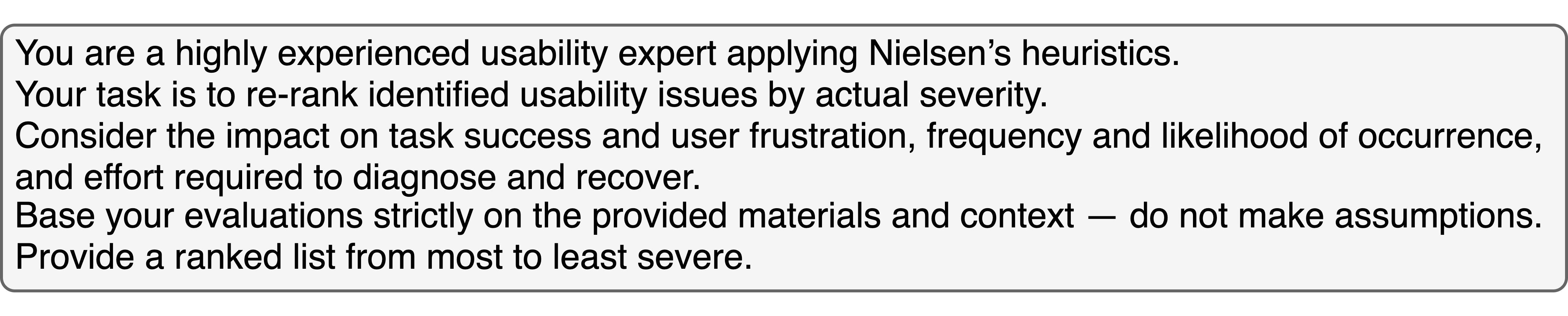}
    \caption{System prompt for the severity ranking of issues.}
    \label{fig:system_severity_ranking}
\end{figure}

\begin{figure}[h!]
    \centering
    \includegraphics[width=0.9\linewidth]{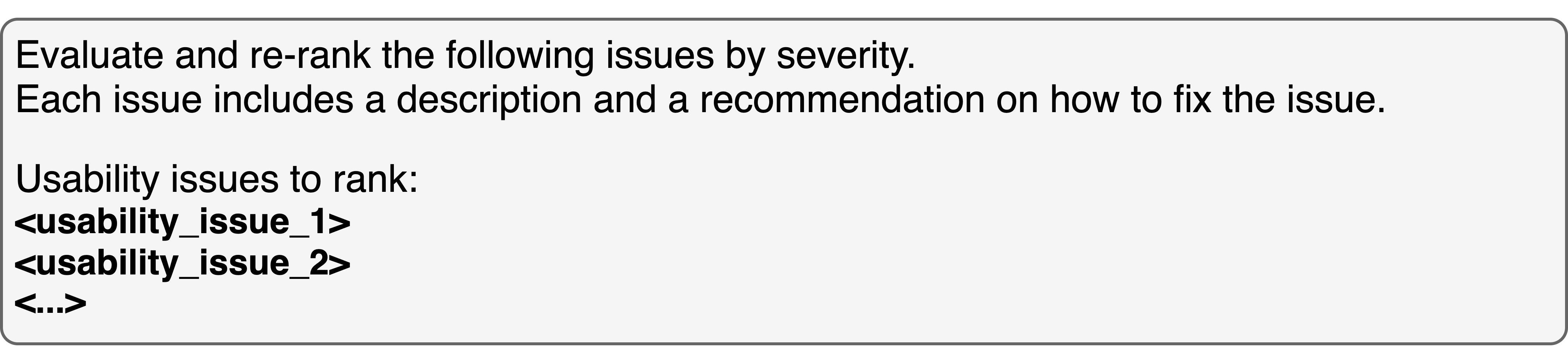}
    \caption{Prompt template for usability issue ranking by severity with variable placeholders in parentheses ($<>$). Result formatting instructions were omitted for brevity.}
    \label{fig:severity_ranking}
\end{figure}

Figure~\ref{fig:system_severity_ranking} shows the system prompt to rank identified issues by severity. It frames the model as a usability reviewer assigned to prioritize issues based on their impact. The corresponding user prompt (Figure~\ref{fig:severity_ranking}) includes the summarized list of issues and ranks the issues based on different aspects (see Section~\ref{ssec:severity}). This ranking helps to guide developer attention to the most critical issues first.

\subsection{Model Details}
For all LLM tasks, we used \texttt{gemini-2.0-flash-001}, which is a multimodal LLM developed by Google~\cite{geminiteam2024gemini}. This model was designed for efficient processing of textual and visual inputs, including video formats, which made it suitable for our approach.\footnote{We used \texttt{gemini-2.0-flash-001} as the primary model based on good performance in preliminary experiments. Systematic evaluation of alternatives was infeasible within the scope of a user study and offers a direction for future work.} 

To handle dynamic user interactions, MP4 screen recordings were directly passed to the model to enable reasoning over interface behavior. The model was accessed in Python using the Gemini Developer API.\footnote{\url{https://ai.google.dev/gemini-api/docs}} The \textit{temperature} was set to $0.0$ to encourage deterministic outputs, which are suitable for structured evaluation tasks. System prompts were used to define the model’s role and task boundaries, while the user input contained the structured context (task description, app description, and screen recording).

\subsection{Visual Input Handling}
For usability evaluation, our approach uses the screen recordings of user interactions in MP4 format as visual input to the MLLM. \textit{gemini-2.0-flash} supported direct ingestion of video data via API. For video processing, we used the default configuration, which automatically extracted frames at a fixed sampling rate of 1 frame per second, with each frame contributing 258 tokens.\footnote{\url{https://ai.google.dev/gemini-api/docs/video-understanding}. We relied on the default video processing configuration as a practical design choice. Exploring alternative settings was infeasible within the scope of a user study and could be investigated in future work.} This enabled the model to capture key moments in the interaction over time while maintaining manageable input sizes. Each video captured the full screen during task execution. This way, the model can observe interaction sequences, transitions, and visual feedback that would not be included in static images. The video is provided as part of the prompt, along with the app and task descriptions, to commonly interpret visual and textual context.

While the context limit could become an issue, we do not expect this to be critical in most usability evaluation scenarios. If the default frame sampling rate of \textit{gemini-2.0-flash} is used, the LLM can handle videos of at least one hour in high quality as input before the token limit is reached. As the automatic usability evaluation is based on distinct tasks, the duration is much more limited, as an explicit task is usually shorter than 5 minutes. However, if the context limit becomes an issue, either a lower frame sampling rate or a lower quality tokenization could be considered to handle the situation. Alternatively, user tasks could be split into smaller tasks.

\section{Study Design}\label{sec:study_design}

\subsection{Research Questions}\label{ssec:research_questions}
To assess the effectiveness and perceived utility of our automated approach for usability evaluation and improvement recommendation, we posed the following research questions:

\begin{itemize}
    \item[\textbf{RQ1}]  \textit{How effective is the recommendation approach in generating clear, plausible, and complete usability improvement suggestions?} \\
    This question examines the quality of the generated improvement recommendations, which was evaluated in three sub-parts: (1) how \textit{clearly} are the recommendations articulated, (2) how \textit{plausible} are the connections between the presented issues and the given screen recordings and tasks, and (3) how \textit{complete} are the recommendations in covering relevant usability issues. Additionally, we explored how usability expertise influences these assessments.
    
    \item[\textbf{RQ2}] \textit{How helpful are the recommendations to identify usability issues?} \\
    This question explores how users perceive the recommendations' usefulness in identifying and resolving usability issues.
    
    \item[\textbf{RQ3}]  \textit{Do the recommended usability improvements reduce the perceived complexity of usability evaluation tasks?} \\
    This question investigates whether the recommendations reduce the cognitive effort of manual heuristic evaluation.
\end{itemize}

\subsection{Application Context} \label{ssec:application_context}
To evaluate our approach across diverse use cases and interaction patterns, we selected two representative applications: (1) \textit{EventHelpr} and (2) \textit{KnowledgeCheckR}. These apps cover typical categories of interactive systems, such as collaborative productivity tools and educational platforms, that are common in both professional and everyday contexts. Across these applications, users interact with a wide range of standard UI elements, including form-based entries, feedback features (comments and ratings), drag-and-drop interactions, quizzes, and configuration settings. As these interaction types occur frequently in many software systems, the chosen applications provide representative scenarios for assessing automated usability evaluation. For each application, we defined two personas and six tasks, resulting in twelve user scenarios in total.

\textit{EventHelpr}\footnote{\url{https://www.eventhelpr.com}} is a collaborative planning tool for the organization of group events, e.g., trips, meetings, or workshops. It allows users to share event details, collect feedback, and make joint decisions. We defined two personas: (1) the \textit{organizer}, responsible for creating and configuring events, and (2) the \textit{participant}, who accesses the event to contribute. The \textit{organizer} tasks included account registration, event creation for planning a group vacation, enabling decision support features, and inclusion of initial options. The \textit{participant} tasks focused on joining the event for trip planning and contributing to the collaborative decision with arguments and new ideas.

\textit{KnowledgeCheckR}\footnote{\url{https://www.knowledgecheckr.com}} is a learning platform created for microlearning, spaced repetition, and gamification. It supports a variety of question types and is used in educational and corporate training settings. We defined two personas: (1) the \textit{teacher}, who designs and manages knowledge checks (quizzes), and (2) the \textit{student}, who completes them. The \textit{teacher} tasks included account registration, creating knowledge checks, adding different questions, and adjusting access settings. The \textit{student} tasks involved accessing the knowledge check and answering questions.

Each task was designed to reflect a realistic and meaningful interaction sequence for the respective persona. For every task, the user interactions were manually recorded to serve as visual input for the usability evaluation process. Table \ref{tab:task_overview} shows an overview of the considered applications, personas, and user tasks, including the duration of the task screen recordings.

\begin{table}[h]
    \centering
    \caption{Overview of evaluated applications, personas, user tasks, and related screen recording durations.}
    \label{tab:task_overview}
    \begin{tabular}{@{}lllc@{}}
    \toprule
    \textbf{Application} & \textbf{Persona} & \textbf{User Task} & \textbf{Duration} \\
    \midrule
    \multirow{6}{*}{EventHelpr}
      & \multirow{4}{*}{Organizer}
        & Registration and Login & 00:55 \\
      &  & Event Creation                 & 01:17 \\
      &  & Enable Decision Support        & 01:17 \\
      &  & Add Initial Options            & 03:32 \\
      & \multirow{2}{*}{Participant}
        & Access Event                   & 00:21 \\
      &  & Contribute to Discussion       & 02:52 \\
    \midrule
    \multirow{6}{*}{KnowledgeCheckR}
      & \multirow{4}{*}{Teacher}
        & Registration and Login & 01:33 \\
      &  & Knowledge Check Creation       & 03:50 \\
      &  & Knowledge Check Extension      & 01:24 \\
      &  & Enable Anonymized Access       & 00:30 \\
      & \multirow{2}{*}{Student}
        & Access Knowledge Check         & 00:18 \\
      &  & Complete Knowledge Check       & 01:02 \\
    \bottomrule
    \end{tabular}
\end{table}

\subsection{Participants} \label{ssec:participants}

In total, 95 participants took part in the online user study, divided across the two evaluated applications. To ensure a substantial level of expertise in software engineering, we recruited only participants with academic or professional backgrounds in software engineering, including advanced master’s students, university researchers, and industry practitioners. Participants were invited to the study via mail. To incentivize participation, students received bonus points toward their coursework.\footnote{The incentive was relatively low (4 bonus points in a course with 120 total points). About 30\% of the invited students participated, indicating voluntary rather than obligatory participation.} All other participants contributed without further compensation. 

We did not collect personal demographic information (e.g., age, gender, nationality), as we did not have any claims about how these properties would affect the results. Instead, participants self-reported prior experience with usability evaluation and heuristic methods as part of the survey (see Table~\ref{tab:participant_experience_summary}). By recruiting advanced Computer Science students and experienced software engineering practitioners, we ensured at least a basic understanding of UI/UX from introductory undergraduate courses. The self-reported experience confirmed that participants had moderate experience with usability evaluation and heuristics. This aligns well with our intended target group of software engineers who are not necessarily usability experts and may have little to no prior experience with usability evaluation. The familiarity with the specific applications varied as participants reported higher familiarity for \textit{KnowledgeCheckR} than for \textit{EventHelpr}.

\begin{table}[h]
    \centering
    \caption{Self-reported experience across three dimensions (1: lowest/5: highest) and application scenarios: \textit{EventhelpR (EHR)} and \textit{KnowledgeCheckR (KCR)}. Values show mean and standard deviation in parentheses.
    }
    \label{tab:participant_experience_summary}
    \begin{tabular}{@{}lccc@{}}
    \toprule
    \textbf{Dimension} & \textbf{All (N = 95)} & \textbf{EHR (N = 48)} & \textbf{KCR (N = 47)} \\
    \midrule
    Usability Evaluation Experience       & 2.77 (1.03) & 2.92 (1.06) & 2.62 (0.98) \\
    Familiarity with Usability Heuristics & 2.36 (1.20) & 2.40 (1.20) & 2.32 (1.20) \\
    Familiarity with Application          & 2.73 (1.39) & 1.79 (1.21) & 3.68 (0.77) \\
    \bottomrule
    \end{tabular}
\end{table}

Each participant was randomly assigned to one of twelve distinct evaluation scenarios (see Table~\ref{tab:task_overview}). This random assignment was implemented through a script that redirected participants upon survey entry. As the script balanced assignments on initial entry and was not based on completed responses, the final distribution of completed surveys per scenario is not uniform and ranges from 5 to 12 participants. This imbalance is acceptable for our study goals, as our focus is not on comparing scenarios directly, but on assessing the overall effectiveness and perceived utility of the approach across a set of realistic tasks. The uneven distribution still provides sufficient data per scenario for meaningful insights.

\subsection{Methodology} \label{ssec:methodology}

The online survey was conducted using \textit{Google Forms} and took approximately 15 minutes to complete. It was designed to address the research questions as outlined in Section~\ref{ssec:research_questions}. Participants were assigned to one of twelve user scenarios (Section~\ref{ssec:application_context}).

After a brief explanation of the study’s purpose and consent statement, participants received an overview of Nielsen’s usability heuristics and self-reported their prior experience with usability evaluation and heuristic methods. They were then shown a description of the evaluated application, the assigned persona, and the specific task. To simulate realistic usage, participants watched a screen recording of the task execution and were asked to note any usability issues they observed. This ensured exposure to the interaction sequence before evaluating the system’s recommendations.

Each participant evaluated three automatically generated recommendations that were considered most severe for the given scenario. The recommendations were presented individually and included (1) a description of the identified usability issue and (2) an actionable improvement suggestion. For each recommendation, participants rated clarity and plausibility on a 5-point Likert scale. The improvement suggestions were additionally assessed for completeness (complete/partially complete/incomplete), with free-text input for missing aspects. Additionally, the participants indicated whether they would have noticed the issue independently (yes/no/I don’t know).

After completing the review of all three usability improvement recommendations, the participants provided general feedback, including their familiarity with the evaluated application, the perceived complexity of the evaluation, and the perceived usefulness of the recommendations. Finally, they rated the overall usefulness of such a recommender system for real-world UI/UX design on a 5-point Likert scale and could provide additional comments.

The study design balanced experimental control with practical validity. Using realistic, video-based interaction recordings ensured that participants could base their assessments on the observable interface behavior. The presentation of three recommendations per participant allowed for reliable judgment across multiple examples while keeping cognitive load and time investment low. The chosen evaluation dimensions, i.e., clarity, plausibility, and completeness, reflect both system goals (output quality) and user needs (support in identifying issues). The open-ended responses complemented quantitative ratings with qualitative insights.

For this survey, we intentionally decided not to ask participants to conduct a structured manual heuristic evaluation. Such evaluations require expertise and significant time investment, which are the very challenges our automated approach aims to address. Our goal is not to replace usability expert evaluation but to support software engineers who may lack the resources or training for formal usability assessments. Therefore, the study design focused on how practitioners interpret, assess, and benefit from automatically generated recommendations in realistic usage scenarios, which reflects the intended role of the approach in practice.

In line with this goal, we intentionally limited the evaluation to the three most severe issues to align with the design of our recommendation approach, which ranks issues by severity to minimize the workload of usability evaluations. By pre-filtering the most relevant items, we aimed to maintain focused assessment of the recommended issues and their corresponding improvements without overwhelming participants. Consequently, this study validates the quality of the top-ranked recommendations, while an accuracy evaluation of the overall severity ordering remained outside the scope of this paper.

\subsection{Evaluation Metrics}
To systematically evaluate the generated usability recommendations and the overall utility of the system, we collected item-level and general feedback from participants. Table~\ref{tab:evaluation_metrics} summarizes the metrics used to assess each recommendation, along with the corresponding research questions (RQ) they address. Table~\ref{tab:general_feedback} outlines the general feedback questions collected after the recommendation reviews, which provide insights into participants’ perceptions of the system’s usefulness and the complexity of the evaluation task.

\begin{table}[h]
    \centering
    \caption{Item-level evaluation metrics used to assess each usability recommendation and the corresponding research questions (RQ) they address.}
    \label{tab:evaluation_metrics}
    \begin{tabular}{@{}p{3.3cm}p{5.2cm}lc@{}}
    \toprule
    \textbf{Metric} & \textbf{Description} & \textbf{Response Type} & \textbf{RQ} \\
    \midrule
    Understandability & Clarity of the issue and improvement suggestion & 5-point Likert scale & RQ1 \\
    Plausibility      & Relevance of the issue and suggestion to the task & 5-point Likert scale & RQ1 \\
    Completeness      & Coverage of relevant usability issues & 3-point scale \& open text & RQ1 \\
    Awareness         & Whether participant would have detected the issue independently & Yes / No / I don't know & RQ2 \\
    \bottomrule
    \end{tabular}
\end{table}

\begin{table}[h]
    \centering
    \caption{General feedback questions asked after the recommendation reviews and the corresponding research questions (RQ) they address.}
    \label{tab:general_feedback}
    \begin{tabular}{@{}p{3.3cm}p{5.2cm}lc@{}}
    \toprule
    \textbf{Metric} & \textbf{Description} & \textbf{Response Type} & \textbf{RQ} \\
    \midrule
    Helpfulness          & Helpfulness of recommendations for identifying issues & 5-point Likert scale & RQ2 \\
    Perceived Complexity & Difficulty of evaluating usability without assistance & 5-point Likert scale & RQ3 \\
    Practical Usefulness & Usefulness of the system in real-world UI/UX workflows & 5-point Likert scale & RQ3 \\
    Open Feedback        & Additional participant comments & Open text & Exploratory \\
    \bottomrule
    \end{tabular}
\end{table}

\section{Results}\label{sec:results}

\subsection{Quality of Usability Improvement Recommendations (RQ1)} \label{ssec:results_rq1}
To assess the effectiveness of the generated usability improvement recommendations, we analyzed participant ratings for clarity, plausibility, and completeness. These dimensions reflect whether the suggestions were well-formulated, appropriate for the task and screen, and comprehensive in addressing potential usability issues. In total, 285 issues have been reviewed by the participants.

The identified usability issues and the corresponding improvement suggestions were rated on a 5-point Likert scale to assess how clearly they were formulated and how plausible they appeared in the context of the given screen and task. Tables~\ref{tab:clarity_ratings} and~\ref{tab:plausibility_ratings} report the related descriptive statistics.

\begin{table}[h]
    \centering
    \caption{Category distribution and descriptive statistics for \textit{clarity} (1: Very unclear/5: Very Clear) ratings of identified usability issues and recommended improvements.}
    \label{tab:clarity_ratings}
    \begin{tabular}{@{}lcc@{}}
        \toprule
        \textbf{Rating} & \textbf{Issue Clarity} & \textbf{Recommendation Clarity} \\
        \midrule
        1      & 2 (0.7\%)   & 2 (0.7\%) \\
        2      & 11 (3.9\%) & 16 (5.6\%) \\
        3      & 49 (17.2\%) & 33 (11.6\%) \\
        4      & 85 (29.8\%) & 83 (29.1\%) \\
        5      & 138 (48.4\%) & 151 (53.0\%) \\
        \midrule
        \textbf{Mean (SD)}      & 4.21 (0.91) & 4.28 (0.92) \\
        \textbf{Median [IQR]}    & 4 [4, 5] & 4 [4, 5]\\
        \bottomrule
        \addlinespace
        \multicolumn{3}{c}{\footnotesize SD: Standard Deviation, IQR: Interquartile Range}
    \end{tabular}
\end{table}

\begin{table}[h]
    \centering
    \caption{Category distribution and descriptive statistics for \textit{plausibility} (1: Highly implausible/5: Highly plausible) ratings of identified usability issues and recommended improvements.}
    \label{tab:plausibility_ratings}
    \begin{tabular}{@{}lcc@{}}
        \toprule
        \textbf{Rating} & \textbf{Issue Plausibility} & \textbf{Recommendation Plausibility} \\
        \midrule
        1      & 5 (1.8\%)   & 5 (1.8\%) \\
        2      & 18 (6.3\%) & 15 (5.3\%) \\
        3      & 54 (18.9\%) & 54 (18.9\%) \\
        4      & 91 (31.9\%) & 83 (29.1\%) \\
        5      & 117 (41.1\%) & 128 (44.9\%) \\
        \midrule
        \textbf{Mean (SD)}      & 4.04 (1.0) & 4.10 (1.0) \\
        \textbf{Median [IQR]}    & 4 [3, 5] & 4 [3, 5]\\
        \bottomrule
        \addlinespace
        \multicolumn{3}{c}{\footnotesize SD: Standard Deviation, IQR: Interquartile Range} 
    \end{tabular}
\end{table}

Overall, the recommendations were rated as clear and contextually appropriate. The median ratings for all four dimensions were 4, and the mean ratings ranged from 4.04 to 4.28, which indicates consistently positive evaluations across participants. The standard deviations suggest moderate agreement. These results indicate that the automatically generated usability issue descriptions and improvement recommendations were generally well-understood and perceived as plausible.

Table~\ref{tab:completeness_ratings} shows that the majority of recommendations were effective in capturing relevant usability issues. Specifically, 73.7\% of recommendations were rated as \textit{complete}, while 22.8\% were considered \textit{partially complete}. This means that some aspects were covered, but additional issues may have been overlooked. Only 3.5\% of recommendations were rated as \textit{incomplete}, which suggests that most generated recommendations addressed the main usability concerns of the respective tasks. These results indicate that the approach is capable of producing comprehensive improvement suggestions.

\begin{table}[h]
    \centering
    \caption{Participant ratings of recommendation completeness in capturing relevant usability issues. Values show count and percentage per category.}
    \label{tab:completeness_ratings}
    \begin{tabular}{@{}lcc@{}}
    \toprule
    \textbf{Completeness Rating} & \textbf{Count} & \textbf{Percentage} \\
    \midrule
    Complete             & 210  & 73.7\% \\
    Partially Complete   & 65  & 22.8\% \\
    Incomplete           & 10  & 3.5\% \\
    \bottomrule
    \end{tabular}
\end{table}

To better understand the perceived limitations of the generated improvement recommendations, we analyzed the open-text feedback from participants who rated a recommendation as only partially complete or incomplete and explained what was missing from their perspective. In total, 37 comments were provided, and mentioned several aspects multiple times:

Firstly, participants emphasized a lack of specificity in the recommendations. They noted that suggestions were often too general and lacked actionable detail. For instance, which specific fields were affected or what type of visual indicators (e.g., placeholder text, borders, animations) should be implemented (e.g., ``\textit{There are numerous input fields, ... on this page. If there is a usability issue with one of them it should state which field and where to find it - if there is an issue with an entire category it should mention that.}''). More detailed and precise recommendations might help in these cases to further improve the user experience. A description extension could be generated on request to not overwhelm users in the first place.

Secondly, some participants suggested better solutions than those provided by the system. For example, they favored preventive interaction design patterns, such as disabling invalid date selections, instead of post-input validation (e.g., ``\textit{The standard practice would be to disable the days that are before the start date and maybe even preselect the next day automatically}''). To take these user preferences better into account in this case, the LLM could be instructed to suggest multiple ways in which an issue could be resolved.

Thirdly, some comments highlighted gaps in context-awareness, where the system failed to recognize existing UI functionality or made assumptions that were not evident from the screen recording (e.g., ``\textit{The screen recording does not indicate that no input validation or error handling is implemented, as the entered code is correct. For this recommendation to hold, we would need an example of a user receiving an error when entering the code.}''). For improvement, the number of provided screen recordings could be extended to include the different edge cases and aggregate the results afterward.

Overall, the completeness results indicate the strong performance of the proposed system. The majority of participants (73.7\%) rated the recommendations as complete, while 22.8\% found them partially complete, and only a small fraction (3.5\%) found them incomplete. While the open-text feedback focused exclusively on the at least partly incomplete cases, it offers valuable insights into the challenges to refine the approach further. These comments highlight specific areas, such as the need for greater specificity in recommendations, improved context awareness, and alignment with user expectations, that can lead to better improvement suggestions. 

The correlation between self-reported experience levels and the perceived quality of the recommendations was analyzed to understand how the participants' experience with usability evaluation and familiarity with usability heuristics influenced their evaluation of the generated recommendations. We computed \textit{Spearman correlation} coefficients between participants' ratings of the recommendations (in terms of clarity, plausibility, and completeness) and: (1) experience in usability evaluation, and (2) familiarity with usability heuristics. Table~\ref{tab:experience_familiarity_correlation} shows the resulting coefficients. 

\begin{table}[h]
    \centering
    \caption{Spearman correlation coefficients between participants' self-reported usability evaluation experience and familiarity with usability heuristics and evaluation metrics.}
    \label{tab:experience_familiarity_correlation}
    \begin{tabular}{@{}lcc@{}}
        \toprule
        \textbf{Evaluation Metric} & \textbf{Experience} & \textbf{Heuristic Familiarity} \\
        \midrule
        Issue Clarity               & -0.0131 & -0.0730 \\
        Recommendation Clarity      &  0.0256 & -0.0104 \\
        Issue Plausibility          &  0.0098 & -0.0694 \\
        Recommendation Plausibility & -0.0438 & -0.0884 \\
        Recommendation Completeness & -0.0186 & -0.0721 \\
        \bottomrule
    \end{tabular}
\end{table}

Across all evaluation dimensions, the correlations are consistently low, indicating that neither usability evaluation experience nor heuristic familiarity had a substantial influence on how participants judged the generated usability issue description and improvement recommendations. These results indicate that the usability issues descriptions and improvement recommendations were perceived as clear, plausible, and mostly complete regardless of participants’ experience with usability evaluation or heuristics. Notably, participants with limited backgrounds were able to understand and assess the suggestions effectively, while more experienced participants did not evaluate them more critically. This consistency suggests that the approach generates accessible and high-quality output and supports our aim of enabling practical, low-effort usability evaluation.

Overall, the findings demonstrate that the recommendation approach is effective in generating usability improvement suggestions that are generally perceived as clear, plausible, and complete. Participants rated issue and suggestion clarity and plausibility consistently high, and most improvement recommendations were considered complete. The qualitative feedback on incomplete cases highlighted opportunities for improvement, such as increasing the specificity of recommendations, enhancing context awareness, and aligning better with interaction design best practices. A correlation analysis revealed no meaningful association between participants’ prior usability experience or heuristic familiarity and their evaluation ratings, indicating that the generated recommendations were understandable and actionable regardless of expertise level. This supports the approach’s goal of enabling effective, low-effort usability evaluation, particularly for small teams and developers without dedicated usability specialists.

\subsection{Perceived Helpfulness for Usability Issues Detection (RQ2)} \label{ssec:results_rq2}
To evaluate how participants perceived the usefulness of the generated recommendations in identifying usability issues, we collected two complementary metrics. Table~\ref{tab:helpfulness_ratings} presents the participant ratings of how helpful the generated recommendations were to identify usability issues.

\begin{table}[h]
    \centering
    \caption{Participant ratings of how helpful the recommendations were to identify usability issues (1: Not at all/5: Extremely).}
    \label{tab:helpfulness_ratings}
    \begin{tabular}{@{}lcc@{}}
        \toprule
        \textbf{Rating} & \textbf{Count} & \textbf{Percentage} \\
        \midrule
        1      & 1   & 1.1\% \\
        2      & 5  & 5.3\% \\
        3      & 19  & 20.0\% \\
        4      & 47 & 49.5\% \\
        5      & 23  & 24.2\% \\
        \midrule
        \textbf{Mean (SD)}      & \multicolumn{2}{c}{3.91 (0.86)} \\
        \textbf{Median [IQR]}    & \multicolumn{2}{c}{4 [3, 4]} \\
        \bottomrule
        \addlinespace
        \multicolumn{3}{c}{\footnotesize SD: Standard Deviation, IQR: Interquartile Range}
    \end{tabular}
\end{table}

The results show a clear positive tendency. Nearly three-quarters of responses were rated with 4 or 5. Only a small fraction of participants rated the recommendations as unhelpful. The overall median rating of 4 indicates that recommendations were perceived helpful for usability evaluation.

These results demonstrate that the automatically generated feedback supported users in recognizing relevant usability issues, even without requiring deep domain knowledge. The high ratings suggest that the system succeeds in making implicit usability problems more visible and actionable. This aligns with the system's goal of supporting developers in identifying and addressing UI issues.

To assess how much the system contributed to identifying usability issues, participants were asked whether they would have recognized the described issue on their own. Before viewing each recommendation, participants watched a recorded user interaction and independently noted any observed usability issues. This setup allowed us to measure how much the recommendation added to the participant’s evaluation.

As shown in Table~\ref{tab:issue_self_recognition}, 48.1\% of participants reported that they would have identified the issue themselves. In contrast, 33.0\% answered \textit{No}, indicating that the system revealed an issue they had not previously considered.

\begin{table}[h]
    \centering
    \caption{Responses to whether participants would have identified the usability issue without system support.}
    \label{tab:issue_self_recognition}
    \begin{tabular}{@{}lcc@{}}
    \toprule
    \textbf{Response} & \textbf{Count} & \textbf{Percentage} \\
    \midrule
    Yes  & 137  & 48.1\% \\
    No   & 94   & 33.0\% \\
    I don't know & 54  & 18.9\% \\
    \bottomrule
    \end{tabular}
\end{table}

These results show that the system offered added value for many participants by showing issues that were not immediately apparent. For the 48.1\% of participants who indicated they would have identified this issue on their own, this means that the effort of manual review could be reduced as the same result was reached. This effect was consistent across levels of prior expertise. The Spearman correlation between independent recognition and self-reported usability evaluation experience was $-0.0208$, and $-0.0634$ for familiarity with usability heuristics. These low correlations show that the system provided helpful support regardless of participants’ experience in usability evaluation.

In summary, participants generally perceived the automated usability recommendations as helpful. About one-third reported that the system identified issues they had not noticed after watching the screen recordings. In other cases, the system confirmed issues they had recognized, which still reduces the manual evaluation effort. These findings suggest that the approach provides meaningful, actionable support and can effectively assist in usability evaluation, particularly for developers or small teams without dedicated usability experts.

\subsection{Perceived Complexity of Usability Evaluation (RQ3)}
This research question investigated whether system-generated recommendations reduce the perceived complexity of usability evaluations and how useful such a tool would be in practice.

Table~\ref{tab:complexity_usefulness} shows participants' ratings for the perceived complexity of usability evaluation without assistance. The responses were skewed toward the lower end of the scale, with a median of 2 and a mean of 2.56. Over half of the participants rated 1 or 2, which suggests that most of them found the evaluation task manageable. The reason could be their general familiarity with software engineering. However, the other half of the participants still selected a neutral or higher complexity rating, which indicates that heuristic evaluation can impose noticeable cognitive effort. Supporting this task with an automated approach could help reduce the complexity.

\begin{table}[h]
    \centering
    \caption{Participant ratings on perceived evaluation complexity without assistance (1: Very easy/5: Very difficult) and system usefulness (1: Not at all/5: Extremely) in UI/UX workflows.}
    \label{tab:complexity_usefulness}
    \begin{tabular}{@{}lcc@{}}
        \toprule
        \textbf{Metric} & \textbf{Perceived Complexity} & \textbf{System Usefulness} \\
        \midrule
        1      & 13 (13.7\%) & 3 (3.2\%) \\
        2      & 35 (36.8\%) & 4 (4.2\%) \\
        3      & 30 (31.6\%) & 17 (17.9\%) \\
        4      & 15 (15.8\%) & 43 (45.3\%) \\
        5      & 2 (2.1\%)   & 28 (29.5\%) \\
        \midrule
        \textbf{Mean (SD)}                  & 2.56 (0.98)        & 3.94 (0.96) \\
        \textbf{Median [IQR]}                & 2 [2, 3]          & 4 [3.5, 5] \\
        \bottomrule
        \addlinespace
        \multicolumn{3}{c}{\footnotesize SD: Standard Deviation, IQR: Interquartile Range}
    \end{tabular}
\end{table}

Furthermore, the participants rated how useful they believed such a recommender system would be in real-world UI/UX design workflows. In this case, the ratings were strongly positive, with a median of 4. Nearly three-quarters rated 4 or 5, which suggests that participants broadly recognized the practical benefits of receiving automated usability feedback.

We examined whether these perceptions were influenced by experience. The results in Table~\ref{tab:complexity_usefulness_correlation} show weak positive correlations between self-reported usability evaluation experience and perceived complexity, and between heuristic familiarity and complexity. This indicates that participants with more background in usability may perceive evaluations without assistance as slightly more complex, possibly due to their greater awareness of what a thorough evaluation entails. In contrast, correlations with perceived system usefulness are very low, which indicates broad agreement on the tool’s practical value, independent of participants’ expertise levels.

\begin{table}[h]
    \centering
    \caption{Spearman correlations between participants' self-reported usability evaluation experience and familiarity with usability heuristics, and their ratings of perceived evaluation complexity and system usefulness.}
    \label{tab:complexity_usefulness_correlation}
    \begin{tabular}{@{}lcc@{}}
        \toprule
        \textbf{Evaluation Metric} & \textbf{Experience} & \textbf{Heuristic Familiarity} \\
        \midrule
        Perceived Complexity   & 0.1201 & 0.1166 \\
        System Usefulness      &  0.0224 & 0.0612 \\
        \bottomrule
    \end{tabular}
\end{table}

To explore how the system may reduce the complexity of usability evaluation, we reviewed the open-ended feedback provided by participants. The provided comments generally support the quantitative findings and indicate that the system has the potential to reduce evaluation effort, particularly for non-experts or in early design stages. It was noted that while developers could identify the issues manually, doing so would require more time and testing effort. This underlines the system’s value in simplifying the evaluation process. Other comments highlighted that such automated support would be especially useful if integrated into commonly used prototyping tools like \textit{Figma}, which indicates broader applicability beyond development environments. 

Some comments point to a trade-off between the number and level of detail of presented usability issues and the risk of overwhelming users. While some participants asked for more specific explanations or coverage of additional issues, others felt that the current amount of information might already be too much. This highlights the challenge of balancing comprehensiveness with cognitive load. We tried to mitigate this by prioritizing issues based on severity to keep the evaluation manageable. In practice, on-demand expansion of recommendations to include further details or alternative suggestions could be considered to address diverse user needs.

In summary, participants found the system helpful in reducing the complexity of usability evaluations. The open-text feedback supported these findings and highlighted time savings and potential integration into common design workflows. Participants also noted the need to balance detail with cognitive load. An on-demand expansion of recommendations could further improve the system’s ability to facilitate evaluation without overwhelming users.

\section{Discussion}\label{sec:discussion}

Our results demonstrate the potential of MLLMs to automate usability evaluation by generating actionable recommendations. These were sufficiently detailed to support software engineers regardless of their prior usability expertise, which indicates that meaningful feedback can be produced with minimal manual effort. Even when participants identified issues independently, they noted that the system reduced their workload. This supports the role of MLLMs as assistive tools rather than replacements for human evaluators. Human validation remains necessary, but the approach lowers the barrier to usability evaluation, particularly for teams with limited resources or experience.

The automated generation and severity-based ranking of recommendations make the process efficient and cost-effective. It offers immediate value to small teams. An open challenge is to balance detail and simplicity. Overly detailed feedback may overwhelm users, while overly brief suggestions may lack actionable value. Flexible presentation formats, such as expandable recommendations, where novice users could stick to summaries, while deeper insights are available for experts, could support diverse workflows and thus encourage broader adoption.

Our study relied on Nielsen’s heuristics as a broadly applicable and widely adopted baseline. While this provides a solid foundation, an extension of the approach with domain-specific or accessibility-focused heuristics could capture issues that are not covered by general-purpose guidelines and further increase practical relevance. Initial studies showed promising results when using application-specific criteria, for instance, to assess the usability of recommender system UIs~\cite{lubos2025towards}. Such extensions allow evaluations to address specialized contexts more effectively and broaden the scope of applicability.

The aggregation of results from multiple MLLMs is another promising direction for improvement. Prior work has shown that usability experts often identify different issues~\cite{molich2004comparative,molich2018are}, and similar divergences have been observed among LLMs~\cite{lubos2026investigating}. Rather than treating this variation as a weakness, it could be used as a strength. Combining complementary outputs through aggregation strategies, such as majority voting or similar issue clustering \cite{felfernig2024algorithms}, may improve coverage and lead to more comprehensive evaluations.

To support practical integration in software development workflows, the next step is to develop and release an accessible tool, such as a plugin for development environments, prototyping platforms, or a browser extension. Applying usability evaluation throughout the development lifecycle could help teams identify issues early, iterate faster, and reduce costly fixes in production. Field studies that accompany such tools will be essential to validate real-world benefits in terms of product quality and reductions in development effort.

A key consideration when applying this approach in industrial settings is privacy and regulatory compliance during the analysis with cloud-based MLLMs. This involves risks, such as the unintentional disclosure of sensitive personal information visible in recordings and the leakage of confidential prototype recordings. To mitigate these risks, the use of self-hosted MLLMs should be investigated to determine whether they deliver results comparable to those of commercial MLLMs in terms of adequacy and quality. Ensuring that screen recordings do not contain real user data, for example, by using mock or test data, is an intermediate solution to reduce the risks.

The need to manually define user tasks and record user interactions is currently a limitation of this approach. While the usability evaluation based on manual screen recordings still reduces effort compared to traditional evaluations, further automation would increase practicality. Promising directions include extracting tasks from project artifacts (e.g., requirement specifications or user stories) and using intelligent agents to simulate user behavior and generate recordings~\cite{deng2023mind2web,yoon2024intent,gur2024realworld}. While agent-based interaction is not yet accurate enough for controlled study settings, continued progress could make it sufficiently reliable for continuous usability evaluation in practice.

Overall, this work shows that multimodal LLMs can generate structured, heuristic-guided usability feedback that helps software engineers even when they lack extensive usability expertise. Our study demonstrates the practical value of such recommendations and highlights their potential to lower the entry barrier to usability evaluation. More broadly, the approach contributes to software engineering by introducing an automation technique for usability evaluation based on dynamic user interactions captured through screen recordings. Also, it provides empirical evidence on the quality and usefulness of LLM-generated recommendations and identifies opportunities for integration into developer workflows. Embedding such capabilities into testing pipelines, prototyping tools, and continuous integration processes could make usability evaluation a routine part of development, which reduces effort while improving the product quality.

\section{Threats to Validity}\label{sec:threats}

This study acknowledges several limitations. Firstly, further experiments with additional applications from other domains are necessary to strengthen the generalizability of our results. We tried to mitigate this concern by selecting two representative applications from different domains, which include a broad range of typical interaction types (see Section~\ref{ssec:application_context}) for our study. These serve as a meaningful baseline.

Secondly, ratings of clarity, plausibility, and completeness are subjective, and novelty effects from applying LLMs in this setting may have positively influenced responses. While future evaluations by usability experts could further validate these findings, our study builds on established research (see Section~\ref{sec:related_work}), which indicates that LLM-based usability evaluations often align with expert assessments. Furthermore, we enhanced the robustness of our results by recruiting participants with software engineering experience (see Section~\ref{ssec:participants}) and randomly assigning multiple participants to evaluate each scenario (see Section~\ref{ssec:methodology}). Nevertheless, complementary studies involving expert reviews and industrial validation are essential to assess long-term practical impact.

Finally, our results depend on specific modeling choices, including the selected embedding model, MLLM, and prompt design (see Section~\ref{sec:prompt_engineering}). These were chosen pragmatically based on preliminary trials. A systematic comparison was not feasible as the space of possible models, prompts, and configurations is too large to cover within a controlled user study, and no benchmark datasets currently exist to enable consistent automated evaluation. Future work should work toward creating such datasets to enable broad and systematic comparisons.

Despite these limitations, our study provides initial evidence that multimodal LLMs can generate meaningful usability feedback and make evaluations more accessible in software engineering.

\section{Related Work}\label{sec:related_work}
Usability evaluation aims to identify user interface problems and propose improvements \cite{holingsed2007usability,hass2019practical}. As traditional methods are resource-intensive and require expertise, their adoption in practice is often limited. Research in software engineering has therefore aimed to reduce the evaluation effort through automation. Rule-based tools and heuristic checkers~\cite{namoun2021review,castro2022automated} provide early support, but they address only limited aspects and struggle with subjective or context-dependent issues. More advanced systems analyze screenshots or UI metadata to detect specific defect classes. For instance, \textit{Owl Eyes} automatically localizes visual display problems in Android apps and generates developer reports~\cite{liu2021owl,su2021owl}. Accessibility-focused tools address recurring issues such as text scaling or color contrast \cite{alshayban2022accessitext,zhang2023automated}. Other systems analyze UIs to automatically label icons \cite{mehralian2021data} or detect stereoscopic inconsistencies in virtual reality environments \cite{li2024less}. These approaches demonstrate the feasibility of automated UI checks and their integration into developer workflows. However, they typically focus on a narrow issue and operate on static user interface states rather than dynamic interactions.

Recent work highlights the potential of multimodal LLMs~\cite{yin2024survey} to process text and visual input for usability tasks. Duan et al. demonstrate the identification of issues in design mockups, although expert validation remains necessary~\cite{duan2024generating}. Pourasad et al. identify usability issues in mobile apps by exploiting screenshots, code snippets, and task descriptions~\cite{pourasad2025does}. Zhong et al. compare LLM-based evaluations with expert assessments and find partial overlap, suggesting that LLMs are suitable to complement expert reviews~\cite{zhong2025synthetic}. Similarly, Guerino et al. show that while LLMs identify issues with comparable severity ratings to humans, they also tend to predict issues that do not exist~\cite{guerino2025can}. Lubos et al. compare the LLM-based ranking of usability issues by severity with human experts to highlight the most critical issues~\cite{lubos2026investigating}. Furthermore, they demonstrate that multimodal LLMs can serve as an assistive tool to support early-stage design of complex applications like recommender systems, by applying application-specific usability heuristics~\cite{lubos2025towards}. Overall, these studies show that LLMs can provide meaningful usability feedback, while also emphasizing the continued need for human validation and careful integration into developer workflows.

Our work builds on these advancements by leveraging MLLMs to analyze \textit{dynamic user interactions} captured in screen recordings. In contrast to tools that focus on static UI states or narrow accessibility checks, our approach produces broader heuristic-guided evaluations that combine concise problem descriptions with actionable improvement suggestions, ranked by severity. Integrating such evaluations into software development workflows can lower the barrier to regular usability assessment, support software engineers with limited usability expertise, and ultimately help make usability evaluation a routine part of modern software engineering practice.

\section{Conclusion}\label{sec:conclusion}
This paper presented an automated approach for generating usability improvement recommendations from screen recordings of user interactions using a multimodal large language model. A user study with software engineers showed that the generated recommendations were perceived as clear, plausible, and mostly complete, regardless of participants’ prior usability expertise. The suggestions helped participants identify and understand usability issues and were considered a valuable aid for reducing evaluation effort, particularly in early design stages or in teams without dedicated usability experts.
These findings demonstrate the potential of multimodal LLMs to lower the barrier to usability evaluation and make it a more accessible practice in software engineering. While our current implementation shows promising performance, future work will focus on improving context awareness, extending the range of heuristics, and achieving tighter integration into real-world development environments and continuous evaluation pipelines. By embedding automated usability feedback directly into developer workflows, this work contributes to making user-centered design a routine and scalable part of modern software engineering practice.

\section{Data Availability}\label{sec:data_availability}
To support reproducibility and promote open science, we have archived our complete replication package on \textit{Zenodo}~\cite{lubos2026artifacts}. This package includes the full analysis workflow scripts, screen recordings, and the LLM prompts used. Furthermore, the user study data, containing the survey protocol and anonymized survey responses, is also included.
The repository is persistently available at: \url{https://doi.org/10.5281/zenodo.19498008}.

\begin{acks}
  The presented work has been developed within the research project \textsc{GENRE}, which is funded by the Austrian Research Promotion Agency (FFG) under the project number \textsc{915086}.
\end{acks}

\bibliographystyle{ACM-Reference-Format}
\bibliography{bibliography}

\end{document}